\documentstyle{article}

\begin{document}
\baselineskip=18pt
\title{A $\hbar$-deformed Virasoro Algebra as Hidden Symmetry of
the Restricted sine-Gordon Model }

\author{
 Bo-yu Hou$^{\dagger}$ 
\hskip 0.5truecm Wen-li Yang$^{\ddag,\dagger}$ \\
\bigskip\\
$^{\ddag}$ CCAST (World Laboratory), P.O.Box 8730, Beijing 100080, China\\
$^{\dagger}$ Institute of Modern Physics, Northwest 
University, Xian 710069, China
\thanks{Mailing address}}
\maketitle

\begin{abstract}
As the Yangian double with center,which is deformed from affine algebra by
the additive loop parameter  $\hbar$ ,we get the commuting relation and
the bosonization of quantum $\hbar$-deformed Virasoro algebra. The
corresponding Miura transformation , associated screening operators and
the BRST charge have been studied.Morever,
we also constructe the bosonization for type I and type II intertwiner vertex
operators.Finally,we show that the commuting relation of these vertex operators
in the case of $p'=r\ \ p=r-1$ and $\hbar =\pi$ actually give the exact
scattering matrix of the Restricted sine-Gordon model.

\end{abstract}

\section{Introduction}
Recently,more and more attention has been paid on the studies of
q-deformation of infinite dimensional algebra---q-deformed affine
algebra$^{[1]}$ ,q-deformed virasoro algebra and W-algebra$^{[3,4]}$ .These
q-deformed algebra would play the same role in the studies of completely
integrable models which are integrablely perturbed from the critical ones
(conformal invariant models) as the non-deformed algebra in the
studies of conformal field theories (CFT) .From the bosonization for
q-deformed affine algebra $U_{q}(\hat{sl_{2}})$ and its vertex
operators, Jimbo et al succeeded in giving the correlation functions
of XXZ model both in the bulk case$^{[7]}$ and the boundary case$^{[8]}$ .
Feigin and Frenkel obtained the q-deformed virasoro algebra,W-algebra,
corresponding ``screening operators" and quantum q-deformed Miura
transformations from the quantization for the q-deformation of the classical
virasoro and W-algebra$^{[4,5]}$. On the other hand,Awata et al also
constructed the quantum q-deformed Virasoro and W-algebra ,associated
``screening operators" and related Miura transformation from the
studies of Macdonald symmetric functions$^{[3]}$.The studies of bosonization
for q-deformed virasoro and W-algebra has been begined in Ref.[2] and Ref.[9].
Actually,Asai,Jimbo et al has constructed the bosonization for vertex
operators of q-deformed W-algebra ($\phi_{\mu}(z)$ in Ref.[9]).

However,there exists another important deformation of infinite dimensional
algebra which also plays an important role in the completely integrable field
theories (In order to comparation with q-deformation , we call it as $\hbar$
-deformation ).This deformation was originated by Drinfeld in studies of
Yangian$^{[1]}$. Simirnov suggested that the Yangian double would be the
dynamical non-abelian symmetry algebra for SU(2)-invariant
Thirring model$^{[16]}$.
Recently progresses has been achieved by Khoroshkin te al $^{[11]}$ and
Iohara et al$^{[12]}$ in the studies of Yangian double  and its central
extension .The bosonization for Yangian double with center and its vertex
operators make it possible to describe the structure of model (local
operators,the asymptotic states etc.)in terms of the representation theory
of Yangian double with center$^{[11]}$.It also ensure that
the correlation functions of SU(2)-invariant Thirring model can
be given explicitly in the integral forms. In this paper,we will given
a new deformation of Virasoro algebra---$\hbar$-deformed Virasoro
algebra(HDVA) which has the similar relation as the Yangian algebra and
the q-deformed affine algebra.When the deformed parameter $\hbar
\longrightarrow 0$, this HDVA will degenerate to the usual Virasoro algebra.
The HDVA is quantum version of some
classical Poisson  algebra which  is constructed by $\hbar$-deformed
Suggware construction of Yangian double with center at the critical point
$^{[4,5]}$( the classical version of this algebra has being studied by our
colleague and the paper is in prepairation).
We also show that when the deformed parameter $\hbar =\pi$, this HDVA
is  the hidden non-abelian dynamical symmetry algebra of the Restricted
sin-Gordon model(This can be seen from the state space$^{[14]}$ and
scattering matrix$^{[15,17]}$). Morever, we constructe
the bosonization for $\hbar$-deformed virasoro algebra and its
type I and type II vertex operators.
The commuting relation for these vertex operators(in the case of
$p'=r\ \ p=r-1$ and $\hbar =\pi$) actually give the exact scattering
matrix of the Restricted sine-Gordon$^{[13,14,15,17]}$.
It is well-known that the similar
phenomena has occured in the studies of braid relations for the
vertex operators (primary fields) of Virasoro algebra in CFT$^{[20]}$.

\section{$\hbar$-deformed virasoro algebra and its bosonization}
Firstly,we give some review of q-deformed virasoro algebra$^{[4]}$.This
deformed algebra 
depends on two parameters p and q,and is generated by current $T_{q}(z)$ with
the following relations
\begin{eqnarray}
& &f_{q}(\frac{w}{z})T_{q}(z)T_{q}(w)-f_{q}(\frac{z}{w})T_{q}(w)T_{q}(z)
=\frac{(1-q)(1-p/q)}{1-p}
(\delta(\frac{w}{zp})-\delta(\frac{wp}{z}))\\
& &f_{q}(x)=\frac{1}{1-x}\frac{(xq;p^{2})(xpq^{-1};p^{2})}
{(xpq;p^{2})(xp^{2}q^{-1};p^{2})}\ \ ,\ \
(z;p)=\prod^{\infty}_{n=0}(1-zp^{n})\nonumber
\end{eqnarray}
It can be shown that the $\hbar$-deformation of affine algebra(or Yangian
double)  can be considered
as the operators scaling limit of q-deformation of affine algebra$^{[2,19]}$
 (but not the scaling limit of bosonic field realization ) .Thus, we give the HDVA as
the operator scaling limit of q-deformation of virasoro algebra in Eq.(1).The
scaling limit is taken as following way
\begin{eqnarray*}
& &z=p^{\frac{-i\beta}{\hbar}}\ \ \, \ \ \ q=p^{-\xi}\ \ ,\ \
f(\beta)=\lim_{p \rightarrow 1}f_{q}(p^{\frac{i\beta}{\hbar}})\\
& &T(\beta)=\lim_{p\rightarrow  1}T_{q}(p^{\frac{-i\beta}{\hbar}})
\end{eqnarray*}
\noindent From direct calculation,we obtain the HDVA which is generated by the
current $T(\beta)$ as the following relation
\begin{eqnarray}
& &f(\beta_{2}-\beta_{1})T(\beta_{1})T(\beta_{2})-f(\beta_{1}-\beta_{2})
T(\beta_{2})T(\beta_{1})=\xi(\xi +1)(\delta(\beta_{1}-\beta_{2}+i\hbar)-
\delta(\beta_{1}-\beta_{2}-i\hbar))\\
& &f(\beta)=\frac{2\Gamma(\frac{i\beta}{2\hbar}+\frac{1}{2}-\frac{\xi}{2})
\Gamma(\frac{i\beta}{2\hbar}+\frac{\xi}{2}+1)}
{\beta\Gamma(\frac{i\beta}{2\hbar}-\frac{\xi}{2})\Gamma(\frac{i\beta}{2\hbar}
+\frac{1}{2}+\frac{\xi}{2})}\nonumber
\end{eqnarray}
\noindent In order to study the bosonization for HDVA ,we will start from the
the quantum Miura transformation associated with HDVA
 in the same way as that of  the studies for
q-deformed W-algebra given by Feigin and Frenkel$^{[4]}$.Let us consider free
bosons $\lambda (t)$ with continuous parameter $t$ ($\in R-{0}$) which satisfy
\begin{eqnarray}
[\lambda(t),\lambda(t')]=\frac{4sh\frac{\hbar t}{2}sh\frac{\hbar\xi t}{2}
sh\frac{\hbar(\xi +1)t}{2}}{tsh\hbar t}\delta(t+t')
\end{eqnarray}
\noindent We also need to define the ``Zero mode" operators P and Q ,which
commute with $\lambda(t)$ and satisfy the relation
\begin{eqnarray}
[P,Q]=-i
\end{eqnarray}
We consider the Fock space $F_{p}$ of Heisenberg algebra defined by Eq.(3)
and Eq.(4) which is generated by the heighest weight vector $v_{p}$.
The heighest weight vector $v_{p}$ satisfies
 \begin{eqnarray}
 \lambda(t)v_{p}=0\ \ \ ,\ \ \ t>0\ \ {\rm and}\ \ Pv_{p}=pv_{p}
 \end{eqnarray}
Now we introduce the field $\Lambda(\beta)$
\begin{eqnarray}
\Lambda(\beta)=:exp\{-\int^{\infty}_{-\infty}\lambda(t)e^{it\beta}dt\}:
\end{eqnarray}
\noindent and give the associated quantum Miura transformation (i.e the
transformation from $\Lambda(\beta)$ to $\hbar$-deformation virasoro algebra
$T(\beta)$ )
\begin{eqnarray}
T(\beta)=\Lambda(\beta+\frac{i\hbar }{2})+\Lambda^{-1}(\beta-
\frac{i\hbar }{2})
\end{eqnarray}
In order to get the commuting relation for bosonic field,we must give a
comment:When one compute the exchange relation of operators,one often
encounter an integral as follows
\begin{eqnarray*}
\int^{\infty}_{0}F(t)dt
\end{eqnarray*}
\noindent which is divergent at $t=0$ .Here we adopt the regularization
given by Jimbo et al$^{[10]}$.Namely, the above integral should be understood
as the contour integral
\begin{eqnarray*}
\int_{c}F(t)\frac{log(-t)}{2i\pi}dt
\end{eqnarray*}
\noindent where the contour C is chosen as in the Ref.[10]. After
straightforward calculation,we show that the bosonic representation for
$\hbar$-deformed virasoro current $T(\beta)$ in Eq.(7) satisfies the
defination relation for HDVA as in Eq.(2).Thus,we constructe the bosonization
for HDVA.

In what follows we will be interested in a particular representions of HDVA
in Fock  space which corresponds to minimal models of CFT$^{[13]}$.This
representations are parameterized by two postive coprime integer numbers $p'$
, $p$ ($p'>p$)  and two type boson $a(t)$ ,$a'(t)$ which are associated with
$\alpha_{+}$ series and $\alpha_{-}$ series respectively
\begin{eqnarray}
& &\xi=\frac{p}{p'-p}\ \ \ ,\ \ \ \alpha_{0}=\sqrt{\xi(1+\xi)}\\
& &\alpha_{+}=-\sqrt{\frac{p'}{p}}=-\sqrt{\frac{(\xi+1)}{\xi}}\ \ \ ,\ \
a(t)=\frac{\lambda(t)}{2sh\frac{t\hbar\xi}{2}}\\
& &\alpha_{-}=\sqrt{\frac{p}{p'}}=\sqrt{\frac{\xi}{(1+\xi)}}\ \ \ ,\ \
a(t)=\frac{\lambda(t)}{2sh\frac{t\hbar (\xi+1)}{2}}
\end{eqnarray}
\noindent We consider the action of HDVA on the Fock space $F_{l,k}$
\begin{eqnarray}
PF_{l,k}=\alpha_{l,k}F_{l,k}\ \ ,\ \ {\rm and}\ \ ,\ \
\alpha_{l,k}=\alpha_{+}l+\alpha_{-}k \ \ ,\ \ 0\leq l \leq p\ \ ,0\leq
k\leq p'
\end{eqnarray}
\noindent as bosonic operator $T(\beta)$ .This action is highly reducible
and we have to throw out some states from Fock module to obtain the
irreducible component. The procedure depends on BRST charge $Q_{\pm}$ which
are defined as follows
\begin{eqnarray}
& &Q_{+}=\oint_{C_{+}}d\beta S^{+}(\beta)f(\beta ,\alpha_{0}P)\ \ ,\ \
Q_{-}=\oint_{C_{-}}d\beta S^{-}(\beta)f'(\beta ,\alpha_{0}P)\\
& &{\rm screening\ \  current\ \ }S^{+}(\beta)=e^{-2i\alpha_{+}Q}:exp\{
-\int^{\infty}_{-\infty}a(t)(e^{\frac{t\hbar}{2}}+e^{-\frac{t\hbar}{2}})
e^{it\beta}dt\}:\\
& &{\rm screening\ \  current\ \ }S^{-}(\beta)=e^{-2i\alpha_{-}Q}:exp\{
-\int^{\infty}_{-\infty}a'(t)(e^{\frac{t\hbar}{2}}+e^{-\frac{t\hbar}{2}})
e^{it\beta}dt\}:\\
& &f(\beta ,\alpha_{0}P)=\frac{sin\pi(\frac{i\beta}{\hbar\xi}
-\frac{1}{2\xi}-\frac{\alpha_{0}P}{\xi})}
{sin\pi(\frac{i\beta}{\hbar\xi}+\frac{1}{2\xi})}\\
& &f'(\beta ,\alpha_{0}P)=\frac{sin\pi(\frac{i\beta}{\hbar (\xi+1)}
+\frac{1}{2(1+\xi)}+\frac{\alpha_{0}P}{1+\xi})}{sin\pi(\frac{i\beta}{\hbar (1+\xi)}
-\frac{1}{2(1+\xi)})}
\end{eqnarray}
\noindent where the integration contours are chosen as follows: the contour
$C_{+}$ enclose the poles $\beta=\frac{i\hbar}{2}-i\hbar\xi n(0\leq n)$ ,the
contour $C_{-}$ enclose the poles $\beta=-\frac{i\hbar}{2}+i\hbar(\xi+1)n
(0\leq n)$ .An important properties of screening charges $Q_{\pm}$ is that
they commute with the HDVA current $T(\beta)$ and have BRST properties
acting on the Fock space $F_{l,k}$
\begin{eqnarray}
Q_{+}^{p}|_{F_{l,k}}=0\ \ ,\ \ Q_{-}^{p'}|_{F_{l,k}}=0
\end{eqnarray}
\noindent As a result, we could have the following complexes
\begin{eqnarray}
& &\ldots \stackrel{X_{-2}=Q^{l}_{+}}{\longrightarrow} F_{2p-l,k}
\stackrel{X_{-1}=Q^{p-l}_{+}}{\longrightarrow} F_{l,k}
\stackrel{X_{0}=Q^{l}_{+}}{\longrightarrow} F_{-l,k}
\stackrel{X_{1}=Q^{p-l}}{\longrightarrow} \ldots\\
& &\ldots \stackrel{Y_{-2}=Q_{-}^{k}}{\longrightarrow} F_{l,2p'-k}
\stackrel{Y_{-1}=Q^{p'-k}_{-}}{\longrightarrow} F_{l,k}
\stackrel{Y_{0}=Q_{-}^{k}}{\longrightarrow} F_{l,-k}
\stackrel{Y_{1}=Q^{p'-k}_{-}}{\longrightarrow} \ldots
\end{eqnarray}
\noindent we assume the following cohomological properties
\begin{eqnarray}
& &Ker X_{j}/Im X_{j-1}=0\ \ ,\ \ if\ \ \ j\neq 0\nonumber\\ 
& &Ker Y_{j}/Im Y_{j-1}=0\ \ ,\ \ if\ \ \ j\neq 0\nonumber\\ 
& &Ker X_{0}=Ker Y_{0}\ \ ,\ \ Im X_{-1}=Im Y_{-1}\nonumber\\
& &{\it L}_{l,k}=Ker X_{0}/Im X_{-1}=Ker Y_{0}/Im Y_{-1}
\end{eqnarray}
\noindent where ${\it L}_{l,k}$ is an irreducible representation of
HDVA.

In order to relate with the Restricted sin-Gordon model, we should take
$\hbar =\pi$ .In the following part of this paper we restrict us to
the case of $\hbar=\pi$
,but the vertex operators ((21)---(24)) and their commuting relations
( (25)----(32)) are easy to generalize to the generic $\hbar$.
If $p=r-1$ and $p'=r$ , the above space ${\it L}_{l,k}$ will be
the space of the Restricted sine-Gordon model $^{[14]}$.

\section{Vertex operators and commuting relations}
It is well-known that the vertex operators of Virasoro algebra ( or of
q-deformed affine algebra) is of great importance in the CFT$^{[13]}$
(or in the solvable lattice model$^{[7]}$). Hence,  we construct the
simplest two type vertex operators: $Z'_{a}(\beta)$ for type I and 
$Z_{a}(\beta)$ for type II
\begin{eqnarray}
& &V_{(1,0)}(\beta)\equiv Z_{+}(\beta)=e^{i\alpha_{+}Q}:exp
\{\int^{-\infty}_{\infty}a(t)e^{it\beta}dt\}:\\
& &V_{(0,1)}(\beta)\equiv Z'_{+}(\beta)=e^{i\alpha_{-}Q}:exp
\{-\int^{-\infty}_{\infty}a'(t)e^{it\beta}dt\}:\\
& &V_{(1,0)}^{(1,0)}(\beta)\equiv Z_{-}(\beta)=\oint_{C_{1}}d\eta Z_{+}(\beta)
S^{+}(\eta)f(\eta-\beta ,\alpha_{0}P)\\
& &V_{(0,1)}^{(0,1)}(\beta)\equiv Z'_{-}(\beta)=\oint_{C_{2}}d\eta
Z_{+}(\beta)S^{-}(\eta)f'(\eta-\beta ,\alpha_{0}P)
\end{eqnarray}
\noindent where the integration contour are chosen as follows: the contour
$C_{1}$ enclose the poles $\eta=\beta+\frac{i\pi}{2}-i\pi\xi n\ \ (0\leq n)$,
the contour $C_{2}$ enclose the poles $\eta=\beta+\frac{i\pi}{2}
+i\pi (1+\xi)n\ \ (0\leq n)$. Because the other vertex operators
$V^{m,n}_{l,k}(\beta)$ (corresponds to $V^{m,n}_{l,k}(z)$ in the Ref.[11])
can be constructed by the symmetric fusion of the simplest two type vertex
operators $Z_{a}(\beta)$ and $Z'_{a}(\beta)$, here we only consider these
two type basical vertex operators.

From the direct caculation, we obtain the commuting relations of vertex
operators(or the braid matrix for the primary fields in CFT$^{[13,20]}$)
\begin{eqnarray}
& &Z_{a}(\beta_{1})Z_{b}(\beta_{2})|_{F_{l,k}}=\sum_{c,d}^{a+b=c+d}
Z_{c}(\beta_{2})Z_{d}(\beta_{1})U\left(\begin{array}{ll}l+a+b&l+d\\l+b&l
\end{array}|\beta_{1}-\beta_{2}\right)|_{F_{l,k}}\\
& &Z'_{a}(\beta_{1})Z'_{b}(\beta_{2})|_{F_{l,k}}=\sum_{c,d}^{a+b=c+d}
Z'_{c}(\beta_{2})Z'_{d}(\beta_{1})U'\left(\begin{array}{ll}k+a+b&k+d\\k+b&k
\end{array}|\beta_{1}-\beta_{2}\right)|_{F_{l,k}}\\
& &Z_{a}(\beta_{1})Z'_{b}(\beta_{2})|_{F_{l,k}}=abctg(\frac{i(\beta_{1}-\beta_{2})}{2}
+\frac{\pi}{4})Z'_{b}(\beta_{2})Z_{a}(\beta_{1})|_{F_{l,k}}
\end{eqnarray}
\begin{eqnarray}
& &U\left(\begin{array}{ll}l\pm2&l\pm1\\l\pm1&l\end{array}|\beta\right)
=\kappa (\beta)\ \ ,\ \
U'\left(\begin{array}{ll}k\pm2&k\pm1\\k\pm1&k\end{array}|\beta\right)
=\kappa '(\beta)\\
& &U\left(\begin{array}{ll}l&l\pm1\\l\pm1&l\end{array}|\beta\right)
=\pm \kappa (\beta)\frac{sin\frac{\pi}{\xi}sin\pi(\frac{i\beta}{\pi\xi}\mp
 \frac{l}{\xi})}{sin\frac{l\pi}{\xi}sin\pi(\frac{i\beta}{\pi\xi}
+ \frac{1}{\xi})}\\    
& &U\left(\begin{array}{ll}l&l\mp1\\l\pm1&l\end{array}|\beta\right)
=\mp \kappa (\beta)\frac{sin\frac{\pi(1\pm l)}{\xi}sin(\frac{i\beta}{\xi})}
{sin\frac{l\pi}{\xi}sin\pi(\frac{i\beta}{\pi\xi}+ \frac{1}{\xi})}\\    
& &U'\left(\begin{array}{ll}k&k\pm1\\k\pm1&k\end{array}|\beta\right)
=\pm \kappa '(\beta)\frac{sin\frac{\pi}{1+\xi}sin\pi(\frac{i\beta}
{\pi(1+\xi)}\pm \frac{k}{1+\xi})}{sin\frac{k\pi}{1+\xi}
sin\pi(\frac{i\beta}{\pi(1+\xi)}
- \frac{1}{1+\xi})}\\    
& &U'\left(\begin{array}{ll}k&k\mp1\\k\pm1&k\end{array}|\beta\right)
=\mp \kappa '(\beta)\frac{sin\frac{i\beta}{1+\xi}sin\pi(\frac{1\pm k}{\xi})}
{sin\frac{k\pi}{1+\xi}sin\pi(\frac{i\beta}{\pi(1+\xi)}-\frac{1}{1+\xi})}\\
& &\kappa (\beta)=exp\{-\int^{\infty}_{0}\frac{sh\frac{t\pi (1+\xi)}{2}
shit\beta}{sh\frac{\pi\xi}{2}ch\frac{t\pi}{2}}\frac{dt}{t}\}\nonumber\\
& &\kappa '(\beta)=exp\{-\int^{\infty}_{0}\frac{sh\frac{t\pi \xi}{2}
shit\beta}{sh\frac{\pi(1+\xi)}{2}ch\frac{t\pi}{2}}\frac{dt}{t}\}
=-\kappa (-\beta)|_{\xi\longrightarrow 1+\xi}\nonumber
\end{eqnarray}
\noindent we also find that
\begin{eqnarray}
[Z_{a}(\beta),Q_{-}]=0\ \ \ ,\ \ [Z'_{a}(\beta),Q_{+}]=0
\end{eqnarray}
\noindent Due to the cohomological properties Eq.(20), the periodic
properties of matrices $U$ and $U'$ with regard to l and k ,and the Eq.(33),
we obtain
\begin{eqnarray}
& &Z_{a}(\beta_{1})Z_{b}(\beta_{2})|_{{\sl L}_{l,k}}=\sum_{c,d}^{a+b=c+d}
Z_{c}(\beta_{2})Z_{d}(\beta_{1})
U\left(\begin{array}{ll}l+a+b&l+d\\l+b&l
\end{array}|\beta_{1}-\beta_{2}\right)
|_{{\sl L}_{l,k}}\\
& &Z'_{a}(\beta_{1})Z'_{b}(\beta_{2})|_{{\sl L}_{l,k}}=\sum_{c,d}^{a+b=c+d}
Z'_{c}(\beta_{2})Z'_{d}(\beta_{1})U'\left(\begin{array}{ll}k+a+b&k+d\\k+b&k
\end{array}|\beta_{1}-\beta_{2}\right)|_{{\sl L}_{l,k}}\\
& &Z_{a}(\beta_{1})Z'_{b}(\beta_{2})|_{{\sl L}_{l,k}}=
ctg(\frac{i(\beta_{1}-\beta_{2})}{2}+\frac{\pi}{4})
Z'_{b}(\beta_{2})Z_{a}(\beta_{1})|_{{\sl L}_{l,k}}
\end{eqnarray}
It is well-known that the Restricted sine-Gordon model is a massive
integrable model which is perturbed from the minimal conformal field model
with center $C=1-\frac{6}{r(r-1)} \ \ (4\leq r)$$^{[14]}$ .We find that when
$p'=r$ and $p=r-1$ (i.e $\xi=r-1$ ) ,besides the space of states are the
same , the braid matrix
$U\left(\begin{array}{ll}l+a+b&l+d\\l+b&l\end{array}|\beta\right)$ is the
exact scattering matrix of the Restricted sine-Gordon model$^{[14,15,17]}$.
 So, we strongly suggest that the $\hbar$-deformed Virasoro algebra defined
 in Eq.(2) is the just hidden symmetry in the Restricted sine-Gordon model.

\section{Conclusion}
In this paper,we obtain the HDVA, its bosonic realization in space 
${\sl L}_{l,k}$ and show that this deformed algebra with the deformed
parameter $\hbar =\pi$ is the hidden symmetry
of the Restricted sine-Gordon model. The bosonization will make it possible
to describe the structure of the Restricted sine-Gordon model in terms of
the representation theory of the HDVA and 
to caculate the correlation functions of the Restricted sine-Gordon model.
The correlation functions would be the scaling limit ($q=p^{-\xi},\ \
\zeta=p^{\frac{i\beta}{\pi}},\ \ r-1=\xi ,p\longrightarrow 1$) of
Lukyanov and Pugain 's in the Ref.[2].We
will present this result in the other paper.Of course,the studies of the
Restricted sine-Gordon model with integrable boundary condition would 
be important and the bosonization of HDVA make it accessible .Some results
abount boundary q-deformed virasoro algebra has been obtained$^{[8,18]}$.

The  $\hbar$-deformation  W-algebra , its bosonic representation and its
vertex operators are also worthy to be investigated.Some results have been
obtained by us which is in prepairation. We expect that these
deformed W-algebra plays the role of symmetry algebra for some integrable
field model.


\begin{thebibliography}{18}

\bibitem{1} Drinfeld,V.G.,Quantum groups,In proceedings of the International
Congress of Mathematicians. pp.798 ,Berkeley, (1987); A new realization
of Yangians and quantized affine algebras,Soviet Math.Dokl. 32(1988), 212.

\bibitem{2} Lukyanov,S., Phys.Letts. B347 (1995) 49; Commn.Math.Phys.
167(1995) 183; Lukyanov,S., and Pugai.Y, Nucl.Phys. B473(1996) 631.

\bibitem{3} Awata,H., Kubo,H., Odake,S. and Shiraishi,J. Comm. Math. Phys.
Vol. 179(1996) 401.

\bibitem{4} Feigin,B. and Frenkel,E., Quantum W-algebras and Elliptic
algebras,q-alg/9508009.

\bibitem{5} Frenkel,.E and Reshetikhin,N. Comm. Math. Phys. Vol.178(1996) 237.

\bibitem{6} Kadeishvili,A.A., Vertex operators for deformed virasoro algebra,
hep-th/9604153.

\bibitem{7} Jimbo,M. and Miwa,T, Analysis of Solvable Lattice Models.
RIMS-981(1994); Jimbo,M., Lashkevich,M., Miwa,T. and Pugai,Y., Lukyanov's
screening operators for the Deformed virasoro algebra, hep-th/9607177.


\bibitem{8} Jimbo,M.,  Kedem,R., Kojima,T., Konno,H. and Miwa,T.,
Nucl.Phys. B441 (1995) 437; Miwa,T. and Weston,.R, Boundary ABF models,
 hep-th/9610094.


\bibitem{9} Asai,Y., Jimbo,M., Miwa.T and Pugai,Y., Bosonization of vertex
operators for $A^{(1)}_{n-1}$ face model, RIMS-1082.

\bibitem{10} Jimbo,M., Konno,H., and Miwa,T., Massless XXZ model and
degeneration of the Elliptic algebra $A_{Q,P}(\hat{sl_{2}})$ ,
RIMS-1105 (1996).

\bibitem{11} Khoroshkin,S., central extension of the Yangian double,
q-alg/9602031; Khoroshkin,S., Lebedev,D. and Pakuliak,S. Traces
of intertwining operators for the Yangian Double, Q-alg/9605039.

\bibitem{12} Iohara,K., and Kohno,M., A central extension of
$DY_{\hbar}(gl_{2})$ and its vertex representations, q-alg/9603032.

\bibitem{13} Felder,G., BRST approach to mininal models, Nucl.Phys.
B317(1989) 215; Felder,G. and LeClair,A., Jour.Mod.Phys. A.7 Suppl. 1A(1992)
239.

\bibitem{14} Eguchi,T. and Yang,S.K., Phys.Letts. B224(1989), 373;
 Phys.Letts. B235(1990) 282.

\bibitem{15} Bernard,D. and LeClair,A., Nucl.Phys. B340(1990) 721;
Comm.Math.Phys. Vol.142(1991) 99; LeClair,A. Phys.Letts.
 B230(1991) 99.

\bibitem{16} Smirnov,F.A., Jour.Mod.Phys. A 7suppl. 1B (1992) 813;
Jour.Mod.Phys. A 7suppl. 1B(1992)839.

\bibitem{17} Reshetikhin,N. and Smirnov,F.A.,
Comm.Math.Phys. Vol.131(1990) 157.

\bibitem{18} Hou,B.Y. and Yang,W.L., Boundary $A^{(1)}_{1}$ face model
(to be published in Letts. of Comm.Theor.Phys.)

\bibitem{19} Hou,B.Y., Shi,K.J., Wang,Y.S. and Yang,W.L., Correlation
functions of SU(2)-Invariant Thirring model with boundary 
(to be published in Jour.Phys.A).

\bibitem{20} Hou,B.Y., Shi,K.J., Wang,P. and Yue,R.H., Nucl.Phys.
B345(1990) 569.



\end{thebibliography}
\end{document}